\documentclass[graybox]{svmult}
   \usepackage{ulem}  
\usepackage{mathptmx}       
\usepackage{helvet}         
\usepackage{courier}        
\usepackage{type1cm}        
\usepackage{makeidx}        
\usepackage{graphicx}       
\usepackage{multicol}       
\usepackage[bottom]{footmisc} 
     \usepackage{ulem}  
\makeindex              

\begin{document}
\newcommand\arcdeg{\mbox{$^\circ$}}%
\newcommand\arcmin{\mbox{$^\prime$}}%
\newcommand\arcsec{\mbox{$^{\prime\prime}$}}%
\newcommand\fd{\mbox{$.\!\!^{\mathrm d}$}}%
\newcommand\fh{\mbox{$.\!\!^{\mathrm h}$}}%
\newcommand\fm{\mbox{$.\!\!^{\mathrm m}$}}%
\newcommand\fs{\mbox{$.\!\!^{\mathrm s}$}}%
\newcommand\fdg{\mbox{$.\!\!^\circ$}}%
\newcommand\micron{\mbox{$\mu$m}}%

\title*{All Things Homunculus}

\author{Nathan Smith}

\institute{Nathan Smith \at Astronomy Department, University of
California, 601 Campbell Hall, Berkeley, CA 94720
\email{nathans@astro.berkeley.edu}}

\maketitle

\abstract{The ``Homunculus'' nebula around Eta Carinae is one of our
  most valuable tools for understanding the extreme nature of episodic
  pre-supernova mass loss in the most massive stars, perhaps even more
  valuable than the historical light curve of $\eta$ Car.  As a young
  nebula that is still in free expansion, it bears the imprint of its
  ejection physics, making it a prototype for understanding the
  bipolar mass loss that is so common in astrophysics.  The high mass
  and kinetic energy of the nebula provide a sobering example of the
  extreme nature of stellar eruptions in massive stars near the
  Eddington limit.  The historical ejection event was observed, and
  current parameters are easily measured due to its impressive flux at
  all wavelengths, so the Homunculus is also a unique laboratory for
  studying rapid dust formation and molecular chemistry, unusual ISM
  abundances, and spectroscopy of dense gas.  Since it is relatively
  nearby and bright and is expanding rapidly, its 3-D geometry,
  kinematics, and detailed structure can be measured accurately,
  providing unusually good quantitative contraints on the physics that
  created these structures.  In this chapter I review the considerable
  recent history of observational and theoretical study of the
  Homunculus nebula, and I provide an up-to-date summary of our
  current understanding, as well as areas that need work.}

   \begin{figure}\center
   \includegraphics[scale=0.73]{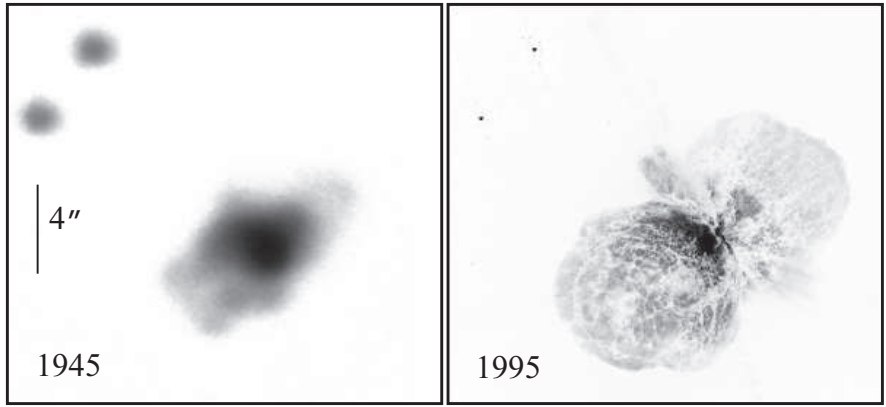}
   \caption{Two visual images of the Homunculus, printed with the same
     spatial scale.  {\it Left}: Gaviola's blue photograph, from an
     original photographic plate that was digitally scanned by the
     author \cite{gav50,sg98}.  {\it Right}: A red-wavelength image
     obtained 50 yr later with the {\it HST} \cite{mors98}.}
   \label{fig:img}
   \end{figure}


\section{EARLY OBSERVATIONS AND PROPER MOTIONS}

Eta Carinae was first recognized as a spatially extended object in
1900--1930 by Innes, van den Bos, and Vo\^{u}te, who noted visual
companion objects less than 2{\arcsec} from the star (see
\cite{adt49,gav50,ring58,sg98}).  Around 1950 the structure was
recognized as nebular by Thackeray and by Gaviola \cite{adt49,gav50}.
Gaviola named it ``the Homunculus'' because on photographic plates it
resembled a small plump man (Fig.~\ref{fig:img}).  The earliest
visible features were mostly in the equatorial parts of the bipolar
configuration. The polar lobes had a diameter of 5--10{\arcsec} in
1900--1940, but the relatively low surface brightness of their outer
extremities made them inconspicuous in the glare of the 3{\arcsec}
central region (see \cite{gav50,dr75}).  During the 1940's the lobes
may have brightened more than the central core region did
\cite{adt53,dvegg52}.  This brightening is probably an important clue
to $\eta$ Car's recovery from the 1843 eruption, but it is difficult
to infer physical information from the scant historical record.

Early measurements of the Homunculus' expansion showed that it was
ejected in the mid-to-late 19th century \cite{gav50,ring58,rdg72}.
The obvious implication, that it was created in the Great Eruption,
was later verified by measurements using data from {\it HST} and with
higher spatial resolution or longer temporal baselines
\cite{cur96a,sg98,mors01}.  Thus we know reliably that the bulk of the
Homunculus nebula was ejected in a relatively short time interval
during the 1840s.  It therefore contains some of our most valuable
information about the Eruption.  This doesn't necessarily mean,
however, that {\it all} the ejecta came from that event. A number of
features appear to have been ejected later, perhaps in the 1890
secondary eruption.  These are discussed later.

\section{THE BRIGHTEST OBJECT IN THE MID-INFRARED SKY}

One of the most important developments in understanding $\eta$ Car was
the recognition that it is extremely luminous at infrared (IR)
wavelengths.  In 1968--1969 Neugebauer \& Westphal discovered a strong
near-IR excess at 1--3~$\mu$m, and then, more important, extremely
luminous mid-IR radiation at 5--20 $\mu$m \cite{nw68,wn69}.  This led
to our present understanding of the energy budget as that of a very
luminous star enshrouded by circumstellar dust, which reprocesses the
UV/visual luminosity into thermal IR radiation
\cite{bejp69a,bejp69b,kd71,rodg71,ks71}.

In subsequent decades, the Homunculus was a favorite target for mid-IR
astronomers because it was bright and spatially extended.  Early
measurements of the spectral energy distribution (SED) established a
range of dust temperatures, 400 to 200~K, a large gas mass of at least
a few M$_{\odot}$, strong and broad silicate emission at 9.7~$\mu$m,
unusually large grains with radii of roughly 1~$\mu$m, and a lack of
strong fine-structure emission lines
\cite{rdg73,rob73,sut74,joy75,ait75,apru75,andr78,mit78}.  The
reasoning for large grains was based mainly on dust temperatures near
the blackbody values for the projected radii of extended features in
mid-IR maps, and on the broad silicate feature \cite{mit86,rob87}, but
this also agreed with the unusually low reddening/extinction ratio at
shorter wavelengths (e.g., \cite{rodg71}). The dusty Homunculus acts
as a calorimeter, providing us with an estimate of the star's total
luminosity $L \; \approx \; 5 \times 10^6 \; L_{\odot}$. This is more
robust than for most other massive stars, because it does not require
a model-dependent spectral extrapolation to UV wavelengths.

The Homunculus showed complex spatial structure in even the earliest
IR drift scans and raster maps that used single-element detectors with
1--2\arcsec\ spatial resolution.  Several authors noted an elongated,
double-peaked, or torus-like structure in the warm inner core, plus a
more extended cooler halo
\cite{hyl79,mit83,chel83,bens85,russ87,allen89}.

In the mid-1980s Hackwell, Gehrz, and Grasdalen made a mid-IR raster
map of the Homunculus, which first clearly delineated the
limb-brightened structure of the two hollow polar lobes (described as
``osculating spheres''), the outer boundary of the mid-IR emitting
structure, and the complex multi-temperature and multi-peaked
structure of the core region \cite{hgg86}.

The advantage of mid-IR imaging of the lobes, compared to reflected
light in visual-wavelength images, is that the thermal dust radiation
in the mid-IR is optically thin, so we can see that the lobes are
hollow and that condensations exist in the obscured core region.
Subsequent imaging with array detectors on 4m-class telescopes
improved upon the spatial resolution (especially in the core), the
sensitivity, and the wavelength coverage, but confirmed the basic
structure\cite{smetal95,sg98,s+98,pol99,morr99,pant00,hony01}\footnote{Note,
  however, that some of the morphological interpretations in these
  studies were contradicted by later observations.}.  So far, the only
high-quality mid-IR polarimetric imaging was presented by Aitken et
al.\ (\cite{ait95}; see also \cite{ba85}), which revealed suggestive
evidence of grain alignment from a possible swept-up ambient magnetic
field, or a strong original field in the star's atmosphere before the
eruption.  Our mid-IR view of the Homunculus improved greatly with the
advent of 8m-class telescopes \cite{s+02,s+03ir,ches05}, because at
long wavelengths like 10--20 $\mu$m, the increased diffraction limit
of larger telescopes is critical.  These modern IR observations will
be discussed along with our modern view of the structure of the nebula
below.

  \begin{figure}\center
  \includegraphics[width=4.5in]{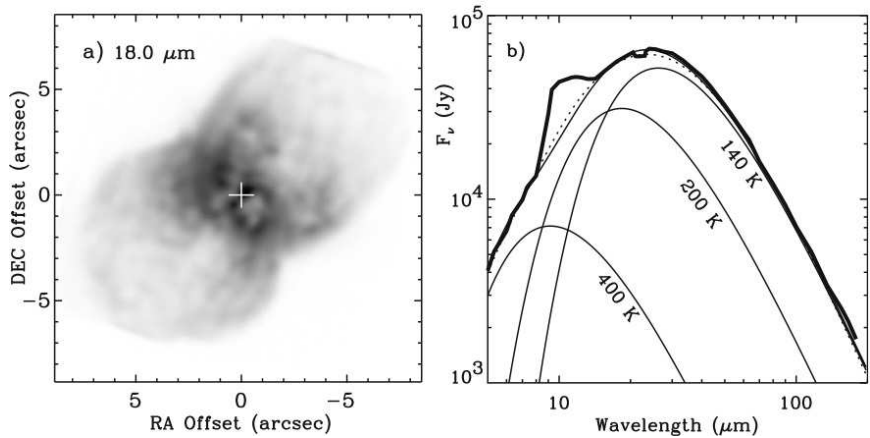}
  \caption{(a) An 18~$\mu$m image of the Homunculus taken with the 6.5-m
     Magellan telescope \cite{s+03ir}. (b) the mid-IR SED of $\eta$ Car
     with the 140 K, 200~K, and warmer component fits used to derive the
     mass in the polar lobes \cite{s+03ir}.}\label{fig:infrared}
  \end{figure}

\section{MASS AND KINETIC ENERGY OF THE HOMUNCULUS}

As noted earlier, the Homunculus is potentially our most valuable
probe of the 19th century eruption and this class of stellar outbursts, 
because with the knowledge from proper motion studies that it was 
indeed ejected in that event, the Homunculus provides us with 
two essential quantities: the ejected mass and its kinetic energy.

The usual way of measuring this mass is to estimate the {\it dust\/}
mass that is needed to produce the IR emission. and then to multiply
by a reasonable gas/dust mass ratio.  With conventional silicate grain
emissivities and gas/dust = 100, the observed 2--12 $\mu$m SED yields
a Homunculus mass of 2--3 $M_{\odot}$, representing the $\sim$200 K
material in the polar lobes \cite{cox95,s+98}.  This is consistent
with the circumstellar extinction \cite{dh97}.  However, dense clumps
which are opaque at visual wavelengths \cite{ebb94,mors98} may contain
additional mass that is under-represented in either type of estimate.
For this reason, it can be problematic to deduce {\it both\/} the
stellar luminosity and the ejecta mass from the same observed IR
luminosity.\footnote{Thus ref.\ \cite{dh97} warned in 1997 that at
  least one of those parameters had presumably been underestimated by
  a then-unknown factor.}

Indeed, the mass estimate of 2--3 $M_{\odot}$ that was adopted for
many years is now known to be too small.  More recent estimates from
dust emission place the ejecta mass (assuming the same gas/dust mass
ratio of 100) at {\it much higher} values of 12.5, 15, or even 20
$M_{\odot}$ \cite{s+03ir,morr99}.  This upward revision followed
chiefly from high-quality measurements at far-IR wavelengths, which
revealed a large mass of dust much cooler than 200 K
\cite{morr99,s+03ir} (Fig.\ \ref{fig:infrared}).

An important distinction must be noted regarding the location of this
cool material.  Morris et al.\ \cite{morr99} fit the far-IR spectrum 
with dust at 110~K, and proposed that it resides in a {\it pre-existing\/} 
massive disk or torus that pinched the waist of the Homunculus when it 
was formed in the Great Eruption.  But Davidson \& Smith \cite{ds00} 
noted that any thermal source at 110~K must cover a much larger 
projected area in order to produce the observed IR flux, and that the 
torus-like structure was already known to be much warmer than 110 K.  
After discovering a thin outer shell of 140~K
dust in high-resolution mid-IR images at 18--25 $\mu$m, Smith et al.
\cite{s+03ir} performed an independent analysis of the same far-IR 
data and found that it could be fit equally well with 140 K dust.  They
suggested instead that the large mass of more than 10 $M_{\odot}$
exists in the polar lobes of the Homunculus, and that there is very
little mass near the equator.  This view had two critical implications: 
(1) the 1840s eruption ejected a large mass in an intrinsically bipolar 
flow, i.e., it was not shaped by pre-existing equatorial material;   
and (2) the $>$10 $M_{\odot}$ of material is moving very fast, with 
a kinetic energy of 10$^{49.6}$--10$^{50}$ erg \cite{s+03ir}.

If the mass and kinetic energy values noted above are incorrect, they
are probably {\it underestimates\/} because the dominant sources of
error tend toward higher mass.  The assumed gas/dust mass ratio of 100
is probably too low given that the ejecta are known to be C and O poor
\cite{kd86}.  Potential optical depth effects, such as shielded cold
grains in dense clumps, can hide additional mass.  Recent observations
at sub-mm wavelengths that are sensitive to the coldest material
indicate 0.3--0.7 $M_{\odot}$ of dust \cite{gom06}, implying 30--70
$M_{\odot}$ of gas if gas/dust = 100 (although some of this emission
may come from a more extended region).  Independent of IR dust
measurements, Smith \& Ferland \cite{sf07} estimated a large {\it gas}
mass of 15--35 $M_{\odot}$ for nebular models that allow H$_2$
molecules to survive near such a luminous star (though values near the
lower end of the range are favored).  So far this is the only estimate
of the ejecta mass that does not rely on an assumed gas/dust ratio.

If $\eta$ Car ejected as much as 20--30 $M_{\odot}$ in a single event
that lasted just a few years, this begins to strain the star's
available mass reservoir if the upper mass limit to stars really is
around 150 $M_{\odot}$, especially if other events like the Great
Eruption have happened once or twice in the past few thousand years.
Such extreme values of the mass and kinetic energy underscore the
profound influence of such high mass-loss events in the evolution of
massive stars \cite{so06,hd94}, not to mention their potential role as
precursors to the most luminous supernova explosions
\cite{s06gy,woo07}.  The fact that they still have no theoretical
explanation makes this one of the most pressing mysteries in
astrophysics, which has not yet received the attention it deserves.

\section{THE BI-POLAR LOBES}

Since the Homunculus is the major product of the 1843 eruption, its
shape, structure, mass, and energy are vital clues for the outburst 
mechanism and the origin of the bipolar shape that is so pervasive 
in astrophysics.  This object has consistently been a favorite target 
of imaging and spectroscopic studies with ever-increasing resolution.  
But it has not yielded its secrets easily, and there has been a 
great deal of controversy over its true structure.  Recent observations 
have helped.

\subsection{Imaging Studies of the Homunculus as a Reflection Nebula}

Although early drawings and photographs showed the polar lobes of the
Homunculus \cite{adt49,gav50,rdg72}, the familiar structure we recognize 
today was not clearly apparent.   Some authors specifically described 
it as ``bipolar'' based on velocities and polarization \cite{ws79,meab87}. 
The first images to show detailed features of the polar lobes and
especially the thin equatorial skirt were ground-based images shown by
Duschl et al.\ \cite{dusch95}.  Later Hubble Space Telescope images 
obtained with a carefully-chosen set of filters and dithered WFPC2 
exposures provided a spectacularly sharp view of the Homunculus 
\cite{mors98}, and became among the most familiar and oft-reproduced 
images made with HST.\footnote{
  [Ed.\ comment:]  Ironically, the lobes are now easily recognizable 
  as such when viewed through an ordinary telescope. Eta Car's 
  visual appearance has changed dramatically in the past 20 years. } 
The bipolar lobes appear remarkably symmetric in these images,
although Morse et al.\ \cite{mors98} remarked that a slight axial
asymmetry can be perceived by simply rotating one of the familiar HST
images by 180\arcdeg. Multiple epochs of images with WFPC2 and HST's
ACS instrument have enabled detailed investigations of the proper
motions and light variations in the
nebula\cite{ebb94,cur96a,s+00,s+04,s+04ph,mors01,jcm06}.

Some of the most notable results from these imaging data are: (1)
intricate, mottled structure on the surfaces of the polar lobes
\cite{mors98}, resembling convective solar granulation or
vegetable-inspired comparisons, (2) the ragged, streaked appearance of
material in the equatorial skirt \cite{dusch95,dh97,mors98}, some of
which seem to connect to more distant nebular features, and (3) the
blue glow or Purple Haze \cite{mors98,s+04ph,zeth99}, a near-UV
flourescent nebulosity in the core of the Homunculus, perhaps
associated with the Little Homunculus or equatorial ejecta \cite{s05}.
HST also revealed remarkable features inside and outside the lobes,
reviewed in other chapters.

Although the Homunculus has structures similar to those seen in
planetary nebulae, it is not a hot photoionized emission-line nebula.
Instead it contains cool low-ionization gas, molecules, and dust
grains.  Its visual spectrum is a complex mixture of reflected
starlight and intrinsic emission from low-ionization species such as
[Fe~II]; the former dominates visual-wavelength radiation at most
locations \cite{meab87,ha92,kd01,s+03lat,zeth99}.  The scattered light
is linearly polarized, with a level of 20--40\% at visual wavelengths
\cite{adt56,adt61,vis67,ws79,schu97,schu99} and 15--30\% in the
near-IR \cite{wal00}.  Even the strong H$\alpha$ emission is mainly
H$\alpha$ from the stellar wind, reflected by grains
\cite{meab93,fal96,s+03lat}.  The surface brightness distribution is
consistent with scattered starlight with moderately large optical
depths \cite{dr75}.  The Homunculus has even been detected as a
Thomson-scattering nebula at hard X-ray wavelengths when the central
source fades\cite{cor04}.

Light scattered by the Homunculus affords us a unique opportunity to
view the star from a range of directions.  The UV/visual spectrum in
the middle of the approaching SE polar lobe gives a nearly pure
(albeit spectrally blurred) reflected spectrum of the star from a
polar direction \cite{kd95,zeth99}. Smith et al.\ \cite{s+03lat}
exploited this fact to derive a 3-D view of the latitude dependence in
the stellar wind of $\eta$ Car, once the 3-D shape and orientation of
the nebula had been determined \cite{kd01,s02}.  Other than the Sun,
this is perhaps the only opportunity in astrophysics where the
spectrum of a star has been viewed from a range of known latitudes.
Likely effects of rapid rotation and a companion star make this of
significant interest.

 \begin{figure}\center           
 \includegraphics[width=4.5in]{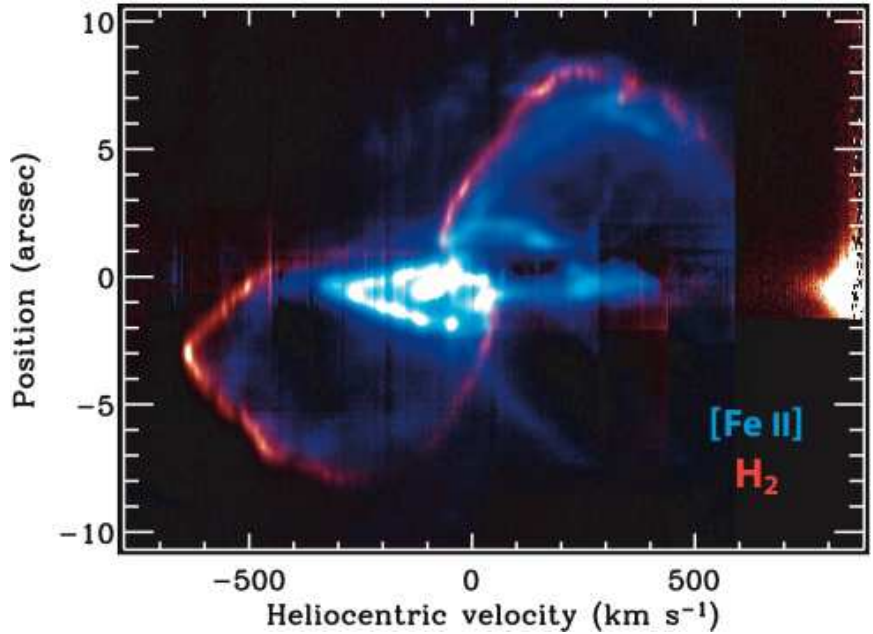} 
 \caption{Long-slit IR spectra of the Homunculus \cite{s06}, showing
   the shape and detailed double-shell structure in H$_2$ (red; thin
   outer shell) and [Fe~{\sc ii}] 1.644~$\mu$m (blue, thicker inner
   shell).  [Fe~{\sc ii}] emission from the Little Homunculus can also
   be seen \cite{s05}.}\label{fig:shape}
 \end{figure}

\subsection{Shape and Orientation of the Polar Lobes}

Structures seen in images are only part of the picture.  Kinematic
information is needed to understand the true 3-D structure.
Pioneering studies by Thackeray, Meaburn, and others combined proper
motions with Doppler velocities to establish that the southeast lobe
is tilted toward us and expanding away from the star with maximum
speeds of roughly 650 km s$^{-1}$, and that the distance to $\eta$ Car
based on expansion is 2.0--2.5 kpc
\cite{adt61,meab87,ha92,meab93,ah93}.  Hillier \& Allen
\cite{ha92,ah93} used long-slit spectra to probe the shape and
expansion properties.  They noted the distinction between narrow lines
like [Ni~{\sc ii}] and [Fe~{\sc ii}] formed in the expanding nebula,
as opposed to dust-scattered components of broad stellar-wind lines or
narrow features arising near the star (see also \cite{meab87,meab93}).
Together these studies reduced the kinematic ambiguities, and
confirmed that the Homunculus is basically a Hubble flow with a
definite age.  Today, with higher spatial resolution studies using
HST/STIS spectra and high-resolution near-IR spectra, we have more
accurate estimates of the inclination $i$ = 42\arcdeg\ (the tilt angle
of the polar axis from our line of sight), the distance of 2.3 kpc
\cite{meab99,kd01,s02,s06}, and the lobe shapes.

For several years, the shape of the polar lobes was discussed in the
context of three alternative models called the ``double bubble'' (two hollow
osculating spheroids), ``bipolar caps,'' and a ``double flask''
\cite{hil97}.  Each found some justification in various aspects of the
imperfect observational data.  The hollow double-bubble model matched
optical and mid-IR images that showed round, limb-brightened lobes
\cite{hgg86,meab93}.  The double cap model emphasized that the polar
regions appear optically thick while the side walls are more
transparent \cite{ah93}.  The flask model had some advantages
concerning polarization and kinematic properties \cite{cur96b}.  In
the end, better data made the distinctions between them largely
irrelevant.

The Space Telescope Imaging Spectrograph (HST/STIS), with long-slit
spectroscopy at high spatial resolution, provided the opportunity to
improve upon the ground-based results.  Davidson et al.\ \cite{kd01}
analyzed some of the same emission and reflection features as Hillier
\& Allen \cite{ha92}, obtaining a much better estimate of the shape
and orientation of the nebula.  A curious discrepancy appeared,
however: polar lobe shapes derived from [Ni~{\sc ii}] emission lines
in STIS data were too narrow and short compared to those seen in
images.  This discrepancy was solved when the first long-slit near-IR
spectra were taken into account \cite{s02}, showing that images in
scattered light corresponded to a thin outer shell seen in H$_2$
emission, while the [Ni~{\sc ii}] emission represented a layer
interior to that.  This double-shell structure is discussed below.

Currently the best estimate of the 3-D shape, structure, and
orientation of the polar lobes is provided by near-IR, long-slit
echelle ($R$=60,000) spectra of H$_2$ emission \cite{s06}.  The H$_2$
traces the dense outer dust shell (see Fig.~\ref{fig:shape}) where
most of the mass resides, and which is seen as the reflection nebula
in images.  The shape and expansion speed are now known with a
precision of roughly 2--3\% at all latitudes in the polar lobes.  The
limiting factor in this study is not the observational precision, but
intrinsic variations and modulations in the lobes themselves, as well
as the assumption that the nebula is axisymmetric.

In other words, the present model for the shape \cite{s06} is as
definitive as the configuration will allow.  Ground-based IR spectra
were even better suited to the task than STIS for three reasons: (1)
their spatial resolution is almost as good as STIS, but the velocity
resolution of the long-slit echelle spectra is much higher; (2) they
trace unique IR molecular hydrogen emission from the outer shell
instead of atomic gas in the inner shell (Fig.~\ref{fig:shape}), and
(3) at near-IR wavelengths one can see all the way through to the far
walls of the polar lobes, mitigating optical-depth effects.  Thus we
now know the 3-D shape of the nebula to high enough precision to
constrain physical models for its formation.

\subsection{Density and Ionization Structure of the Polar Lobes}

In addition to the overall shape, recent observations from large
ground-based telescopes and {\it HST} have significantly advanced
our understanding of the detailed density and ionization structure of
the Homunculus.  Until recently, the walls of the polar lobes were
generally assumed to have a thickness of $\sim$10\% of their radius
\cite{hgg86,s+98}, and aside from the clumpy structure in {\it HST}
images, not much more was known about their detailed structure.

Now we know that the walls actually have a well-defined double-shell
structure (Fig.~\ref{fig:shape}; \cite{s06}), where most of the mass
is in a denser and geometrically thinner outer molecular shell (traced
most clearly by near-IR H$_2$ emission), and about 10\% of the mass in
a geometrically thicker but optically thinner inner shell of
partially-ionized atomic gas (traced by near-IR and optical emission
species [Fe~{\sc ii}] and [Ni~{\sc ii}]).  This double-shell structure
was first proposed based on H$_2$ and [Fe~{\sc ii}] structures seen in
the first available long-slit near-IR spectra of the Homunculus
\cite{s02}, although subsequent long-slit spectra of the same lines
with higher resolution provided a much more definitive picture
\cite{s06}.

The double-shell structure is verified by 8$-$25~$\mu$m imaging on
large ground based telescopes with sub-arcsecond resolution
\cite{s+03ir}, revealing a thin outer shell of cool dust with a color
temperature of 140~K, and a thicker inner shell of warmer dust at
200~K in the polar lobes (Fig.~\ref{fig:infrared}).  The thicker
and warmer inner shell dominated previous mid-IR imaging at
shorter wavelengths near 10~$\mu$m;  Smith et al.\ \cite{s+98} noted
that the warm dust emitting at 10--12~$\mu$m was $\sim$10\% less
extended than the scattered light seen at visual and near-IR
wavelengths, but did not articulate the nature of the double-shell
structure.  Fits to the IR SED imply that the outer 140~K shell
contains about 11~$M_{\odot}$ of material, while the inner 200~K shell
has only 1.5~$M_{\odot}$ \cite{s+03ir}.\footnote{  
  Pantin \& Le Mignant \cite{pant00} proposed an ``onion-like'' 
  structure for the polar lobes, with a smaller bubble inside 
  a larger one.  This structure was not verified in mid-IR images 
  with higher resolution, and their proposed structure was not 
  the double-shell structure descibed here. }

The effects of this double-shell structure are also clearly seen in UV
echelle spectra obtained with STIS, tracing absorption along our line
of sight to the central star.  These spectra show narrow absorption
lines from Fe~{\sc i} and diatomic molecules in the thin outer shell
at $-$530 km s$^{-1}$, and a more clumpy and broader distribution of
absorption lines from singly-ionized metals at slower blueshifted
speeds \cite{gull05,gull06,niel05,ver05}.  For the thin outer shell,
these same studies suggest that the gas temperature is $\sim$760~K,
significantly warmer than the 140~K dust in the same location
\cite{s+03ir}.

The double-shell structure is explained quantitatively in model
calculations of the nebula's ionization structure \cite{sf07}.  The
inner shell is a warm (neutral-H) photodissociation region where
metals like Fe and Ni are singly ionized by strong Balmer continuum
radiation, while the outer thin and cool molecular shell exists where
a thin layer of H$_2$ has absorbed much of the radiation in the
Lyman-Werner bands.  The required neutral H densities are roughly
10$^{5.5}$ cm$^{-3}$ in the inner shell (or more if the material is
clumpy), and (0.5--1)$\times$10$^7$ cm$^{-3}$ in the outer shell in
order to allow H$_2$ to survive and for Fe$^+$ to recombine to Fe$^0$
\cite{sf07}.  UV absorption features also suggest densities of 10$^7$
cm$^{-3}$ in the thin outer molecular shell along our line of sight
\cite{gull05}.  With densities this high, the observed volume of the
cool shell would indicate high masses for the Homunculus of $\sim$20
$M_{\odot}$ \cite{s06,sf07}.

What gives rise to this double-shell structure?  On the one hand, the
presence of near-IR H$_2$ and [Fe~{\sc ii}] emission features remind
one of warm shocked gas \cite{sd01}, in two zones behind the forward
and reverse shocks, respectively.  This would be somewhat misleading,
however: comparing radiative energy heating to shock heating indicates
that radiative heating dominates the energy balance by a factor of 250
or more \cite{sf07}.  Thus, the near-IR lines in the Homunculus do not
trace current shock excitation as in a supernova remnant, for example,
but instead trace different zones of ionization in a dense photodissociation
region \cite{sf07}.  In near-IR lines of H$_2$ and [Fe~{\sc ii}], this
double-shell structure of the Homunculus is qualitatively identical to
that seen in the planetary nebula M 2-9 \cite{HL94,sbg05}, where UV
excitation is implied by the H$_2$ line ratios.  If the density
structure was imprinted by a shock, that shock interaction must have
happened long ago \cite{s06}.  Further work on this is needed.

Another unsolved issue is the origin of the clumpy, mottled structure
apparent on the surfaces of the polar lobes.  The true nature of the
dark lanes vs.\ bright clumps, and of the obvious ``hole'' at the
polar axis in the SE lobe have been topics of debate because different
wavelengths or different techniques (imaging, spectroscopy,
polarimetry) yield conflicting answers
\cite{mors98,s+98,s+03ir,s06,schu99}.  Recent spectra of H$_2$ have
shown that the ``hole'' really is an absence of material \cite{s06},
and IFU spectra of [Fe~{\sc ii}] emission confirm its presence in the
inner shell of the SE lobe as well as the receding NW polar lobe
\cite{teo08}.  The nature of the cells and dark lanes remains an open
question, however, and the answer potentially holds important clues to
whether they originated in Rayleigh-Taylor instabilities, thermal
instabilities, convective cells, or porosity
\cite{dh97,s+98,s06,stan04}.

\subsection{What Caused the Bipolar Shape?}

Most theoretical work on the Homunculus has attempted to answer this
question.  The answer, which is not yet clear, is likely to be
intimately connected to the cause of the Great Eruption, the star's
behavior before and after, and the potential roles of rotation and/or
binarity in the system.

In some early models, the bipolar form of the Homunculus was said to 
arise from the interaction of winds of differing speed and density, 
shaping the outflow after ejection \cite{f+95,db98,lang99}.
These borrowed from models for the ring and bipolar nebula of SN~1987A 
and planetary nebulae \cite{BL93,ma95,fm94,f99}.  The key requirement 
was a pre-existing, slow and dense equatorial disk or torus to pinch the
waist of the bipolar structure.  Observations, however, then showed 
that the equatorial skirt is no older than the polar lobes and some 
parts are younger \cite{sg98,mors01,kd01}.  We now know that there is 
not enough material near the equator to shape the overall morphology;
instead, most of the mass is located at high latitudes in the polar 
lobes \cite{s06}.

The next models to explain the morphology assumed that ``intrinsic'' 
shaping mechanisms had occurred in the ejection process.  Acknowledging 
the apparent absence of any pre-existing 
``doughnut,'' Frank et al.\ \cite{f+98} suggested that 
the wind during the Great Eruption was fast and highly
aspherical.  This hypothesis was extended to the shaping of the Little
Homunculus as well \cite{gonz04a,gonz04b}.  While they yield bipolar
nebulae resembling the Homunculus, these hydrodynamic models did not
explain why the eruption wind was aspherical to begin with.  

Smith et al.\ \cite{s+03lat} showed that $\eta$ Car's present-day wind 
is in fact bipolar with more mass and higher speeds at the poles, 
and noted that if the present-day wind were given a mass-loss rate 1000 
times stronger, it would look like the ballistically expanding 
Homunculus.  IR observations \cite{s02,s06,s+03ir} also showed
that the shape of the lobes approximately agrees with what one  
expects for the latitudinal variation in escape speed on the
surface of a rotating star, that more mass is located at the
poles of the Homunculus, and that the thickness of the lobes compared
to their radius is the same as the duration of the eruption divided
by its age.  The intrinsic bipolar mass ejection must presumably 
then be related to either rotation or to a companion star (or 
rotation induced by a companion star).  In the context of aspherical 
line-driven winds, Owocki and coworkers \cite{stan97,stan96,stan98} 
proposed that gravity darkening and velocity-dependent forces on 
a rotating star inhibit equatorial mass loss and {\it enhance the 
mass loss toward the poles}.  Thus it was suggested that 
the Homunculus lobes arose from a bipolar continuum-driven wind
with a higher polar mass-loss rate because of the latitudinal
variation in escape speed  \cite{stan03,stan05,stan97,md01,do02}. 
This would require an enhanced wind for a brief time during the 
eruption, ejected directly from the surface of a rotating luminous star.  

An intrinsically bipolar wind from a rotating star has several advantages 
over a pre-existing torus that pinches the waist, but there was still 
no explanation for the origin of the thin equatorial skirt described 
in \S \ref{equatorial} below.  Smith \& Townsend
\cite{st07} noted that since disk-inhibition mechanims arise because
of velocity-dependent forces in a line-driven wind \cite{stan96}, a
wind-compressed disk \cite{bc93} could still be created in a 
continuum-driven wind.  In that case the disk is formed by hyperbolic 
orbits crossing the equator after being launched from the surface of a
rotating star, while the enhanced polar mass flux arises because of
gravity darkening.   Smith \& Townsend showed that one can thereby 
account for both the shape of the polar lobes and a thin fast 
equatorial ejecta skirt.

In all the models mentioned so far, mass ejection by the primary star
is the principle agent for shaping the nebula. (This includes a
scenario where the rotating primary has increased angular momentum
through interaction with a nearby companion \cite{s+03lat}).  Binary
models have also been proposed to explain the bipolar shape.  Soker
\cite{sok01,sok04,sok05,sok07} advocates an analytic accretion model,
where a close companion star in a binary system accretes matter from
the primary wind through Bondi-Hoyle accretion, and then launches
bipolar jets or collimated winds that shape the polar lobes.
Alternatively, the idea of a binary merger (i.e. a merger of a close
binary in a hierarchical triple system) has been suggested more than
once to explain the Great Eruption and the origin of the bipolar
nebula \cite{jsg89,iben99,mp06}, but it is difficult to understand the
formation of the thin disk or the occurrence of previous giant
eruptions in this type of scenario.

\section{DUST, MOLECULES AND CHEMISTRY}   

By virtue of its phenomenal IR luminosity, the Homunculus has long
been appreciated as a unique laboratory for the study of dust
formation and survival in harsh conditions.  The grains in the
Homunculus are unusual compared to grains in the ISM and AGB stars in
two important ways:

{\bf (1)} They seem to be unusually large, $a \, \sim \, 1$~$\mu$m 
or more, based on the following clues.  The grains are near equilibrium 
blackbody temperatures \cite{sf07,s+03ir,s+98,hgg86,mit86,mit83}, their 
UV-to-IR extinction ratio is extraordinarily gray 
\cite{rodg71,rs67,andr78,whit94,ha92,hil01},
and they scatter efficiently even at $\sim$2~$\mu$m \cite{sg00}.  
(Polarization properties of the nebula indicate that some small 
grains are also present \cite{schu99}.)  

{\bf (2)} They may have unusual composition, probably as a
result of C and O depletion in the ejecta.  While $\eta$
Car shows what appears to be a strong 9.7~$\mu$m silicate feature,
it is unusually broad and shifted to longer wavelengths.  This has
prompted suggestions that alternative grain minerology such as
corrundum (Al$_2$O$_3$) and olivine (MgFeSiO$_4$) might be present
\cite{mit78,ches05}.  In particular, Chesneau et al.\ \cite{ches05}
demonstrated that a combination of olivine and corrundum can account
for the observed shape of the spatially dependent ``silicate'' feature
in the core of the Homunculus.  The unusual composition of the grains
may suggest that current mass estimates of the Homunculus of 10--15
$M_{\odot}$ may be underestimates \cite{dekoter05} (see above for
other reasons why the mass estimates are conservative).

Our understanding of how these large and unusual grains nucleated
rapidly and survived in the outflow of the Great Eruption is far from
complete, and the puzzle deserves further investigation both
theoretically and observationally.  An additional mystery concerns the
fact that gas-phase Fe has basically solar abundance in such a dusty
region as the Homunculus \cite{sf07}.  Since we know when the material
was ejected and when the dust formed from the historical record, and
$\eta$ Car is bright enough in the IR to provide high-quality data today,
the Homunculus is a rare and valuable laboratory to address these
questions.

In addition to unusual dust grains, the Homunculus harbors a
surprisingly large mass of molecular gas, less than 0.1 pc from one of
the most luminous hot supergiants known (i.e., $\eta$ Car itself).
Molecules were first discovered there via near-IR H$_2$ emission lines
\cite{sd01,s02}, and were later shown to reside in the cool outer skin
of each polar lobe \cite{s02,s06}.  Given the presence of molecular
hydrogen in a presumably nitrogen-rich environment, one might expect
species like NH$_3$, which indeed was the first polyatomic molecule
detected in the Homunculus \cite{sNH3}.  This was also the first
detection of a polyatomic molecule around any LBV.  UV absorption
studies with STIS have revealed diatomic molecules such as CH and OH
in the thin outer shell \cite{ver05}.  These UV absorption
observations and radio studies have found no evidence for CO,
presumably because most CNO in $\eta$ Car's ejecta is N rather than C
and O.  While the outer ejecta and the inner Wiegelt Knots are known
to be N-rich from their atomic emission-line spectra
\cite{kd86,kd95,sm04}, the presence of ammonia gives our first
reliable verification that the Homunculus too is N-rich
\cite{sNH3}. The formation and survival of these molecules is far from
understood, and more theoretical work on the molecular chemistry of
the Homunculus would be useful \cite{fer05}.

\section{THE LITTLE HOMUNCULUS}

The Little Homunculus (LH) is a smaller bipolar nebula nested inside
the main Homunculus, oriented along the same bipolar axis \cite{bish03}.  
It is a relatively new addition to $\eta$ Car's family of known 
morphological features.  The LH was briefly called 
the ``integral-sign filament'' (referring to the shape of its emission 
lines in {\it HST\/}/STIS spectrograms) and the ``Matryoshka Nebula'' 
\cite{gull99,bish01}.

Its discovery by Ishibashi et al.\/ was based on its Doppler-shift
morphology in visual-wavelength spectra \cite{bish03}; bright
reflection nebulosity and extinction in the Homunculus make its
structure very difficult to perceive in visual-wavelength images.  The
LH's 3-D structure is best seen in near-IR emission lines of [Fe~{\sc
  ii}], because these lines are bright and they can penetrate the
foreground dust screen \cite{s02,s05}. This structure is clearly
evident in Figure~\ref{fig:shape}.  The LH spans 4--5\arcsec\ along
the polar axis of the Homunculus \cite{s05}.  It does not align with
any features in visual scattered light, nor does it match the inner IR
torus \cite{s05}.  Since it was not detected at 10--20 $\mu$m, it
probably does not contain a substantial amount of dust \cite{s+03ir}.

It does, however, match up quite well with the ``Purple Haze''
\cite{s+04ph,mors98} and the inner radio continuum nebula
\cite{dun97}.  Smith \cite{s05} proposed that periodic and
direction-dependent UV illumination of the LH is responsible for the
extended radio continuum nebula and its associated changes through the
5.5~yr cycle.  This could offer a powerful test of models for the 5.5
yr variability.  Because it is seen in the radio continuum, the LH
apparently absorbs whatever last remaining ionizing photons escape the
dense wind of the central system.

Most indications suggest that the LH was ejected during $\eta$ Car's
1890 eruption.  The LH has an expansion speed of 300 km s$^{-1}$ at
the poles and slower values at lower latitudes, with an expansion
speed of $\sim$140 km s$^{-1}$ along our line of sight \cite{s05}.
Thus, the LH is responsible for the $-$146 km s$^{-1}$ absorption
feature seen in high resolution UV spectra \cite{gull05}.  Walborn \&
Liller \cite{wl77} noted blueshifted absorption features around 200 km
s$^{-1}$ in the spectrum observed during the 1890 event, making this
association seem plausible.  Nebular kinematics \cite{s05} and proper
motions in STIS spectra \cite{bish05} both imply an ejection date
around 1910, but this assumes constant linear expansion.  The LH's
polar expansion speed of 300 km s$^{-1}$ is much slower than the
500--600 km s$^{-1}$ stellar wind of $\eta$ Car that pushes behind it,
and Smith \cite{s05} showed that the LH would indeed be accelerated to
its present value if it had been ejected in 1890, and that this
acceleration leads us to infer an incorrect and more recent ejection
date.  The same may be true for the kinematics of the Weigelt knots
\cite{s+04}.  Smith \cite{s05} proposed that the Weigelt knots are
simply the equatorial pinched waist of the LH, analogous to the IR
torus of the larger Homunculus (see below). See also the chapter by
Weigelt and Strauss on the inner ejecta in this volume.

Estimates of densities in the LH range from a few times 10$^4$
cm$^{-3}$ \cite{s02} to 10$^6$ or even 10$^7$ cm$^{-3}$ \cite{bish03}.
Combining the observed 3-D volume of the LH with an uncertain
ionization fraction, the mass is estimated as 0.1 $M_{\odot}$
\cite{s05}, which is in rough agreement with other estimates based on
upper limits to the dust mass and the kinematics.  The kinetic energy
of the 1890 event is then roughly 10$^{46.9}$ erg \cite{s05}.  Given
that the mass and kinetic energy of the 1890 event were orders of
magnitude less than the Great Eruption \cite{s+03ir}, the fact that
the two mass ejections shared the same basic geometry is thought
provoking, to say the least.  From a similar study of infrared
[Fe~{\sc ii}] emission from P~Cygni's shell, Smith \& Hartigan
\cite{sh06} concluded that the 1600 AD giant eruption of P Cygni was
nearly identical to the 1890 lesser eruption of $\eta$ Car in terms of
its ejected mass, speed, and kinetic energy, although it was not
bipolar. (See the chapter on the outer ejecta and other LBV-associated 
nebulae by Weis, this volume.)

\section{EQUATORIAL STRUCTURES}
\label{equatorial}

\subsection{The Equatorial Skirt}

While bipolar lobes often occur in planetary nebulae and other
astronomical objects, the ragged debris disk in $\eta$ Carinae's
equatorial plane is an uncommon and somewhat mysterious sight
\cite{dh97}. This equatorial ``skirt'' or ``debris'', as it is
sometimes called to distinguish it from a smooth (and especially a
Keplerian) disk, was first seen in high-resolution ground-based images
\cite{dusch95}.  Images with {\it HST} showed more detail in the
radial streaks and other features \cite{ebb94,dh97,kd97,mors98}.
Images alone were inadequate, however, because some ``equatorial
features'' later proved illusory.\footnote{ For instance, a bright
  fan-shaped structure 4\arcsec\ northwest of the star was often cited
  in the 1990s as an especially obvious equatorial feature.  But
  Doppler velocities and IR measurements show that it is really a
  less-obscured patch of the NW polar lobe, despite appearances
  \cite{kd01,s+98,s+03ir}. } The geometrical thinness, planar form,
and velocity structure of the skirt were especially apparent in HST
spectral data \cite{kd01}, and Doppler velocities remain necessary to
establish whether a feature is equatorial or not.

Though prominent in scattered-light images with brightness comparable
to the polar lobes, the equatorial skirt is not seen clearly in
thermal-IR maps \cite{hgg86,s+98,s+03ir,pol99}, suggesting that it
contains little mass or that it is not heated efficiently.  If the
dust were cooler and substantially massive, however, one would expect
the equatorial features to become more prominent at longer wavelengths
or to cause severe extinction, but they are as hard to see at
25~$\mu$m as at 10~$\mu$m \cite{s+03ir}.  Smith et al.\ \cite{s+03ir}
have discussed the illumination of the equatorial ejecta in detail.

The age of the equatorial ejecta inspired a lively debate which is
still not entirely settled \cite{cur96a,sg98,cur99,mors01}. Proper
motions at low spatial resolution with a time baseline of 50 years
yielded an ejection date around 1885 for the equatorial features,
suggesting that they came from the 1890 event \cite{sg98}.  Gaviola
noted hints toward the same conclusion \cite{gav50,ring58}.  Doppler
velocities indicate post-1860 material in the skirt
\cite{dh97,kd97,kd01,s02,s05} and in the roughly-equatorial Weigelt
Knots closer to the star \cite{kd97,dor04,s+04,w95}.  Proper motions
measured in high-resolution {\it HST} data \cite{cur99,mors01},
however, suggested that the skirt features were older, perhaps coeval
with the Homunculus.  Subsequent spectra showed emission lines in the
equatorial material with {\it both} ages \cite{kd01}.  Thus, the
resolution of the disagreement may be that the early blue plates
(1945) and color slides (1972) were contaminated by line emission from
younger diffuse gas originating in the 1890 event, whereas the small
clumps measured in {\it HST} images were dust condensations tracing
most of the mass.  At least one of the features overlapped with the
``Purple Haze'' \cite{s+04ph}, a region of line emission that may be
partly due to the younger Little Homunculus.

Only two models so far have attempted to explain the simultaneous
origin of the polar lobes and this equatorial skirt, and in both cases
the origin of the asymmetry was an inherently aspherical ejection by
the star.  Matt \& Balick \cite{mb04} presented a magnetohydrodynamic
model for the present-day bipolar wind and disk, and conjectured that
if this occurred at much higher mass-loss rate it may also explain the
Homunculus.  Smith \& Townsend \cite{st07} presented a model of
ejection from the surface of a rotating star that would produce a
structure like the skirt via a compressed disk, along with the polar
lobes.  This type of mechanism is intuitively appealing because the
required ``splashing'' at the equator may partly account for the
ragged streaked appearance of the skirt \cite{dh97,kd97}.

Some prominent features in the skirt share apparent connections to
more distant and perplexing features in the outer nebulosity.  For
instance, the bright equatorial structure located $\sim$4\arcsec\
north-east of the star seems to connect to a nebular feature,
sometimes termed the NN Condensation, north-east ``jet'', or ``NN
jet'' \cite{walborn73,walborn78,meab93b}.  This structure extends
radially in the same direction, driving an apparent bow-shock
structure in the outer ejecta that is seen in nebular emission-line
images \cite{meab87,meab93b,mors98}.  Similarly, the streaked features
in the skirt located west of the star appear to connect to the
S~Condensation.  These perceived connections may be due to either
common hydrodynamic flows or illusions caused by beams of starlight
escaping in preferred directions, or both.  These collimated
protrusions\footnote{One hesitates to call them ``jets'', as they lack
  evidence of being steady collimated flows, but appear to have
  resulted from episodic mass ejections instead (e.g., \cite{meab87}.}
in the equatorial plane have few parallels in astrophysical objects.
These and additional structures in the outer ejecta are discussed
further in the chapter by Weis.

  \begin{figure}\center
  \includegraphics[width=3.5in]{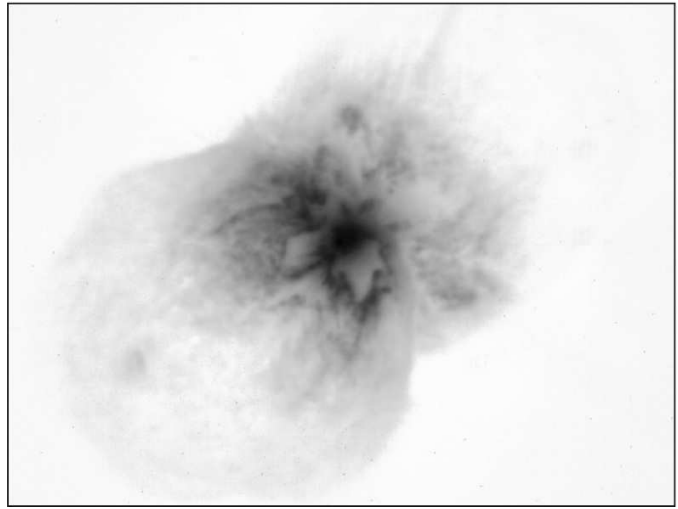}
  \caption{A near-IR adaptive-optics image of the core of the
    Homunculus obtained with the VLT.  This figure was prepared by the
    author from a $K$-band image kindly provided by O.\ Chesneau (see
    \cite{ches05}).}\label{fig:vlt}
  \end{figure}

\subsection{The Obscured Inner Torus}

As noted earlier, IR imaging during the 1970s--1990s revealed a
complex, multi-peaked structure in the core of the Homunculus.
Several near-IR and mid-IR studies with 2--4-m class telescopes, each
of which improved on pevious work, unveiled these inner structures at
high resolution, most often interpreted as some variation of a dusty
torus \cite{hgg86,smetal95,rg95,fal96,s+98,morr99,pol99,sg00,hony01}.
Access to mid-IR imaging on 6--8 m class telescopes improved our view
of the torus \cite{s+02}, and we now have stunning adaptive optics
images in the near-IR (Fig.~\ref{fig:vlt}) obtained with the VLT
\cite{ches05}.

Like the larger equatorial skirt, this compact and complex structure
has sparked interesting debates.  Morris et al.\ \cite{morr99}
proposed that this structure was a pre-existing 15 $M_{\odot}$ torus,
including the 110~K dust that domainted the far-IR flux, and was the
agent responsible for pinching the waist of the Homunculus.  Davidson
\& Smith \cite{ds00} argued against this interpretation, and pointed
out that the torus was actually quite warm.  Hony et al.\
\cite{hony01} proposed that the inner structure was a pair of polar
rings like those around SN~1987A, and that the polar axis of the
binary system had precessed by 37--58\arcdeg\ caused by a close
interaction of a triple system.  Smith et al.\ \cite{s+02} presented
higher resolution images showing that these rings were an artifact of
lower spatial resolution, resembling a fragmented torus instead.
Chesneau et al.\ \cite{ches05} presented even higher-resolution images
showing surprisingly complex structure (Fig.~\ref{fig:vlt}), and
proposed that these unprecedented features were part of an inner
bipolar nebula that they named the ``butterfly nebula''.

Examination of narrow H$_2$ velocities in near-IR long-slit echelle
spectra showed that the dusty features in question were in the
mid-plane of the Homunculus where the two polar lobes meet at the
pinched waist of the nebula, and were not related to polar features of
the Little Homunculus \cite{s05,s06}.  The expected dust temperatures
at this location agree with the observed warm color temperatures of
the dust of 250--400 K \cite{s+03ir}.  The strange and irregularly
corrugated structures seen in the VLT images \cite{ches05} probably
result from the strong post-eruption wind of $\eta$ Car pushing out
the relatively low density regions in between clumps in the torus
\cite{s+02}.  In some cases, perhaps these protrusions have broken all
the way through in the mid plane, allowing the stars visual light to
escape to large radii, producing radial streaks seen in images of
reflected light.  This provides only a partial explanation for outer
features like the NN jet and the S Condensation, which still evade our
understanding. See relevant chapters this volume by Weis and by
Weigelt and Strauss.

The clumpy torus seen in near- and mid-IR images has no counterpart in
visual-wavelength images in scattered light, but it can be seen by
virtue of its emission-line variability caused by ionization changes
during the 5.5 yr cycle \cite{s+00}, and it can be recognized by its
Doppler shifted narrow line emission in long-slit STIS spectra of the
inner Homunculus \cite{gull01}.

\section{SUMMARY: FAST FACTS}

The Homunculus of $\eta$ Car is a spectacular and complex object that
is a rich laboratory for studying many phenomena associated with ISM
processes and the physics of stellar mass loss.  It is 
in its scattered starlight at
visual and near-IR wavelengths, allowing us to see the star from
multiple directions \cite{s+03lat}, but it also has intrinsic emission
features that tell us about the kinematics and excitation of the
nebula \cite{ha92,kd01,s06,sf07}.  Its intrinsic spectrum is that of a
low-ionization warm photodissociation region, but with peculiar gas
and dust abundances.  Extended emission has been detected at all
wavelengths from scattered X-rays \cite{cor04}, the UV \cite{s+04},
visual and near-IR wavelengths, thermal-IR, and even low-level radio
continuum emission from the polar lobes \cite{dun97}.  It is rivaled
by few other objects in the sky, but its complexity often defies the
simplest interpretation.  Our understanding of the Homunculus will
no-doubt continue to improve, but here I list current best estimates
for several observed properties of the nebula:

\begin{itemize}

\item{Proper motions measured with post-refurbishment {\it HST} images
   give an ejection date for the Homunculus of 1847 ($\pm$4--6 yr)
   \cite{mors01}.  This is consistent with the 1843 eruption peak
   within the uncertainty, but hints that there may have been some
   post-ejection acceleration.}

\item{The total IR luminosity integrated from 2--200~$\mu$m is
   4.3$\times$10$^6$ L$_{\odot}$ \cite{s+03ir}.  The total stellar
   luminosity is 10--20\% higher than this, since that much UV/visual
   light escapes.}

\item{The total mass of the Homunculus (gas + dust) is at least 12--15
   $M_{\odot}$ \cite{s+03ir,morr99,sf07}, but could be substantially
   more as noted above.  The total kinetic energy is then at least
   10$^{49.7}$ erg \cite{s+03ir}.}

\item{The shape of the polar lobes is traced best by near-IR H$_2$
   emission \cite{s06} and is shown in Figure~\ref{fig:shape}.  Smith
   \cite{s06} also gives the expansion speed, mass, and kinetic
   energy as a function of latitude.  Most of the mass and kinetic
   energy is at high latitudes.}

\item{The heliocentric distance to the Homunculus, derived from its
   proper motion and observed Doppler shifts assuming axisymmetry, is
   2.3$\pm$0.05 kpc \cite{meab99,kd01,s02,s06}.}

\item{The inclination angle of the Homunculus (the angle that the
   polar axis is tilted from the line-of-sight) is
   $i$=42\arcdeg$\pm$1\arcdeg\ \cite{kd01,s06}.}

\item{The systemic velocity of $\eta$ Car is $-$8.1($\pm$1) km
   s$^{-1}$ heliocentric, or $-$19.7($\pm$1) km s$^{-1}$ LSR,
   measured from narrow H$_2$ lines \cite{s04}.}

\item{The Homunculus has a well-defined double-shell structure.  The
   thin outer shell has about 90\% of the total mass, is primarily
   molecular gas, Fe is neutral, the particle density is $\sim$10$^7$
   cm$^{-3}$, and the dust temperature is about 140~K.  The inner
   shell is thicker and less massive, the dust is warmer at about
   200~K, metals like Fe and Ni are singly ionized, and H is mostly
   neutral.}

\item{The dust grains that dominate thermal emission and excitation
   are large, with $a\simeq$1~$\mu$m.  Judging by the 10~$\mu$m
   emission feature, the chemical makup of the dust appears to be
   unusual compared to other evolved stars, with significant amounts
   of corrundum and olivine.}

\item{The detection of NH$_3$ and lack of CO verifies that the
   neutral/molecular Homunculus is N-rich like the outer ejecta and
   Weigelt knots \cite{sNH3}.  The molecules that have been detected
   so far are H$_2$ \cite{s02}, CH and OH \cite{ver05}, and NH$_3$
   \cite{sNH3}.  Verner et al.\ suggest that CH$^+$, CH$_2^+$,
   CH$_2$, and NH may also be detectable.  One might also expect to
   find N$_2$H$^+$ \cite{sNH3}.  In other words, the Homunculus is a
   valuable laboratory for studying N-rich molecular chemistry.}

\item{Most of the dense clumpy structures in the equatorial skirt seen
   in optical images were ejected at the same time as the polar
   lobes.  There is also younger material from the 1890 eruption
   intermixed with these equatorial ejecta, seen mainly in emission
   lines.  The equatorial skirt is {\it not} a pre-existing disk that
   pinched the waist of the nebula.}

\item{Infrared images show a bright IR torus inside the Homunculus
   that is not seen in visual-wavelength images.  It is bright in the
   IR because it contains some of the warmest dust in the Homunculus,
   where the thin walls of the two polar lobes meet at the equator.
   The mass in this torus is not nearly enough to have pinched the
   waist of the Homunculus.}

\end{itemize}

\section{THE FUTURE FOR THE HOMUNCULUS}

Eta Carinae is among the most bizarre extended objects
in the sky, with some of the most spectacularly complex and puzzling
circumstellar structures known.  It is wise to keep this in
perspective, however, remembering that many of the reasons the
Homunculus seems so peculiar arise {\it because it is so young}.  It
is young enough that it is still in free expansion, and so the
observed geometry still bears the imprint of the initial ejection --
before its signature is erased by deceleration from the surrounding
medium.  This makes $\eta$ Car unique among massive stars, and this
is why it is such a valuable tool for constraining the physics of
episodic mass loss from massive stars.  It is therefore instructive to
end by briefly considering the future fate of the Homunculus:

In another 500--1000 years, the Homunculus will have expanded to
$\sim$5 times its current size, and will plow into the outer ejecta.
Its very clean bipolar and thin disk geometry will probably be diluted
or erased.  It will no longer obscure the central star (and will
therefore be harder to observe), and it will not be one of the
brightest 10--20~$\mu$m objects in the sky because the dust will
absorb a smaller fraction of the total luminosity and will be much
cooler. At lower densities it will probably be largely ionized (if not
by $\eta$ Car, then by the remaining $\sim$60 O-type stars in the Carina
nebula), and the thin outer molecular shell will probably not survive.

In other words, it will probably appear as a fairly ill-defined,
hollow ellipsoidal ionized gas shell with unclear kinematics, much
like $\eta$ Car's current outer ejecta from a previous eruption or the
nebula around AG~Car \cite{stahl,s97,voo00}.  In context, then, the
Homunculus is perhaps not so bizarre after all, since shell nebulae
with masses of $\sim$10 $M_{\odot}$ are not unusual for LBVs with
luminosities above 10$^6$ L$_{\odot}$ \cite{so06}.  This epitaph
underscores how lucky we are to observe the Homunculus in our
lifetimes with available technology at an optimal time when $\tau$=1
throughout much of the nebula.  Until the distant (or not so distant?)
future when fuel has been exhausted in the star's core, the Homunculus
will continue to disperse, leaving the star unobscured when it
explodes...unless another giant eruption happens first.

\begin{acknowledgement}

 I am indebted to numerous people with whom I have collaborated on
 studies of the Homunculus and discussed related topics, especially
 Kris Davidson, Gary Ferland, Bob Gehrz, John Hillier, Jon Morse, 
 Stan Owocki, and Rich Townsend.

\end{acknowledgement}


\end{document}